\documentclass[journal]{IEEEtran}

\usepackage{url}
\usepackage{./bs_macros}
\usepackage{caption}
\usepackage{cancel}
\usepackage{subcaption}
\usepackage{balance}

\usepackage{algorithm}
\usepackage[noend]{algpseudocode}

\graphicspath{{./}}

\theoremstyle{definition}

\newtheorem{remark}{Remark}

\makeatletter
\patchcmd{\@maketitle}
{\addvspace{0.5\baselineskip}\egroup}
{\addvspace{-1.3\baselineskip}\egroup}
{}
{}
\makeatother

\begin{document}

\title{A Class of Doubly Stochastic Shift Operators for Random Graph Signals and their Boundedness}

%
%
%
%
%

\author{
	
Bruno Scalzo Dees, \IEEEmembership{Student Member, IEEE}, Ljubi$\check{\text{s}}$a Stankovi\'c, \IEEEmembership{Fellow, IEEE}, Milo$\check{\text{s}}$ Dakovi\'c, \IEEEmembership{Member, IEEE}, Anthony G. Constantinides, \IEEEmembership{Life Fellow, IEEE}, Danilo P. Mandic, \IEEEmembership{Fellow, IEEE} 
	
\thanks{B. Scalzo Dees, A. G. Constantinides and D. P. Mandic are with the Department of Electrical and Electronic Engineering, Imperial College London, London SW7 2AZ, U.K., e-mail: \{bruno.scalzo-dees12, a.constantinides, d.mandic\}@imperial.ac.uk.}
\thanks{L. Stankovi\'c and M. Dakovi\'c are with the with the Faculty of Electrical Engineering, University of Montenegro, D$\check{\text{z}}$ord$\check{\text{z}}$a Va$\check{\text{s}}$ingtona bb, 81000 Podgorica, Montenegro, e-mail: \{ljubisa, milos\}@ucg.ac.me.}


}


\maketitle

\begin{abstract}
A class of doubly stochastic graph shift operators (GSO) is proposed, which is shown to exhibit: (i) lower and upper $L_{2}$-boundedness for locally stationary random graph signals; (ii) $L_{2}$-isometry for \textit{i.i.d.} random graph signals with the asymptotic increase in the incoming neighbourhood size of vertices; and (iii) preservation of the mean of any graph signal. These properties are obtained through a statistical consistency analysis of the graph shift, and by exploiting the dual role of the doubly stochastic GSO as a Markov (diffusion) matrix and as an unbiased expectation operator. Practical utility of the class of doubly stochastic GSOs is demonstrated in a real-world multi-sensor signal filtering setting.
\end{abstract}


\IEEEpeerreviewmaketitle

\vspace{-0.5cm}

\section{Introduction}

Given the rapidly increasing availability of data recorded on irregular domains, it would be extremely advantageous to analyse such unstructured data as signals on graphs and thus benefit from the ability of graphs to incorporate domain-specific knowledge. This has motivated the developments in the emerging field of Graph Signal Processing \cite{Sandryhaila2013,Shuman2013,Sandryhaila2015,Ortega2018,Stankovic2019_1,Stankovic2019_2}, and has spurred the introduction of the graph counterparts of many classical signal processing algorithms.

One such direction is that of the system (or filter) on a graph, which was recently considered in \cite{Sandryhaila2013,Ribeiro2017,Ribeiro2017_2,Eldar2017,Stankovic2019_2}. In classical signal processing, a system is typically a linear operator that maps an input signal to another (output) signal. However, while the signal \textit{shift operator} (unit time delay) is the lynchpin in discrete-time linear systems, its definition on graphs is not obvious due to the rich underlying connectivity structure. Indeed, the shift of a random graph signal can be viewed as the diffusion of a signal sample from the considered vertex along all edges connected to this vertex. Therefore, to effectively employ a system which operates on random signals acquired on graphs, it is necessary to rigorously consider the statistical properties of such \textit{graph shift operators} (GSOs).

Existing GSOs typically take the form of the graph adjacency or Laplacian matrices, which are in general neither bounded nor isometric operators. Without this property the repeated application of a shift to a graph signal can significantly alter or distort the spectral content of the graph signal, thus making it difficult to design and understand the filter frequency response as in classical signal processing \cite{Gavili2017}. Therefore, for rigour it may not only be desirable but also necessary to preserve the signal energy ($L_{2}$-norm) over shifts. For instance, strict-sense stationary graph signals are defined to be statistically invariant to graph shifts, therefore, the isometry property is necessary in order to perform statistical operations on stationary graph signals in a mathematically rigorous and tractable manner \cite{Girault2015,Girault2015_3,Girault2015_2}.

Several isometric shift operators have been recently proposed which satisfy the desirable isometry property \cite{Girault2015,Girault2015_3,Girault2015_2,Gavili2017}. These graph shift operators are constructed as a diagonal matrix, with the entries defined as the eigenvalues of the adjacency or Laplacian matrix cast onto a unit circle, thus preserving in this way the isometry property. However, there remain issues that need to be addressed prior to a more widespread application of existing isometric shift operators: 
\begin{enumerate}[label=(\roman*)]
	\item The isometry property is not guaranteed when the graph is directed as the eigenvalues of the adjacency or weight matrix can be complex-valued \cite{Girault2016};
	\item The eigenvalues used to construct the isometric GSO, which relate to the $L_{2}$-norm of graph edge weights, are sensitive to low-probability (outlier) edge weights;
	\item Important localization properties of the graph are lost by defining the GSO as a diagonal matrix \cite{Vandergheynst2017}.
\end{enumerate}
Instead, it may be more suitable to employ graph shift operators which achieve tight boundedness, or even the isometry property with respect to metrics other than the $L_{2}$-norm, for a wide range of random graph signals.

To this end, we investigate the statistical and boundedness properties of a class of doubly stochastic graph shift operators, which are shown to exhibit the following desirable properties:
\begin{enumerate}[label=(\roman*)]
	\item $L_{2}$-norm of locally stationary random graph signals is upper and lower bounded over shifts;
	\item $L_{2}$-norm isometry is attained for \textit{i.i.d.} random graph signals with an asymptotic increase in the vertex incoming neighbourhood size;
	\item The mean of the graph signal is preserved over shifts.
\end{enumerate}
These boundedness properties are derived by employing the left-stochastic property (each column sums up to unity), which allows for a Markovian (diffusion) interpretation of the graph shift, and the right-stochastic property (each row sums up to unity), for the GSO to be viewed as an unbiased expectation operator. In this way, the examination of the boundedness properties reduces to performing a statistical consistency analysis of the graph shift. Practical utility of this class of GSOs is demonstrated through a physically meaningful and intuitive real-world example of geographically distributed estimation of multi-sensor temperature measurements.

\pagebreak

\section{Doubly Stochastic Graph Shift Operators}


\subsection{Preliminaries}

The signal domains considered in this work are \textit{graphs}, whereby a graph, denoted by $\mathcal{G} = \{\mathcal{V},\mathcal{E}\}$, is defined as a set of $N$ vertices, $\mathcal{V} = \{1,2,...,N\}$, which are connected by a set of edges, $\mathcal{E} \subset \mathcal{V} \times \mathcal{V}$. The existence of an edge going from vertex $m$ to vertex $n$ is designated by $(m, n) \in \mathcal{E}$. 

The \textit{incoming} neighbourhood of a vertex $m$, denoted by $\mathcal{V}_{m} \subset \mathcal{V}$, is the subset of vertices, $n$, for which $(n,m) \in \mathcal{E}$. The size of this neighbourhood, $\mathcal{V}_{m}$, is denoted by $N_{m}$.

The strength of connectivity of an $N$-vertex graph can be represented by the \textit{weighted adjacency matrix}, $\W \in \domR^{N \times N}$, with its entries defined as 
\begin{equation} \label{eq:weight_matrix}
W_{mn} \begin{cases} 
> 0, & (n,m) \in \mathcal{E},\\
=0, & (n,m) \notin \mathcal{E},
\end{cases}
\end{equation}
whereby the amplitude of the entries conveys the \textit{relative} importance of the vertex connections. Regarding the directionality of vertex connections, a graph is undirected if each edge, $(m, n) \in \mathcal{E}$, has its counterpart, $(n, m) \in \mathcal{E}$, such that $\W = \W^{\Trans}$. For generality, in this work we consider directed graphs, for which this symmetry property does not hold. 



\subsection{Random signals on a graph}

\label{sec:random_graphs}

With each vertex, $n \in \mathcal{V}$, we can associate a real-valued random variable, $x_{n} \in \domR$. Upon considering all vertices in $\mathcal{V}$, a random signal on a graph is denoted by $\x \in \domR^{N}$ and is said to be wide-sense stationary (WSS) if and only if its first- and second-order moments are invariant under the application of a graph shift \cite{Girault2015,Girault2015_3,Girault2015_2}. By defining the GSO as $\S \in \domR^{N \times N}$, the conditions for graph wide-sense stationarity are given by $\expect{\x} \! = \! \expect{\S\x}$ and $\expect{\x\x^{\Trans}} \! = \! \expect{\S\x\x^{\Trans}\S^{\Trans}}$. Notice that in this way the graph ensemble mean, $\expect{\x}$, is not a constant but a vertex-varying signal. Various Graph Signal Processing applications have been investigated based on this definition of the WSS graph signal \cite{Vandergheynst2017,Ribeiro2017_2,Ribeiro2019}.

Despite the desirable mathematical tractability of WSS graph signals, they do not appropriately model the smoothness of the nonstationarity in the vertex-domain inherent to real-world graph signals. Consequently, a class of \textit{locally} stationary graph signals was introduced in \cite{Girault2017,Girault2017_2}, whereby the statistical properties of the vertex signals within a neighbourhood are assumed to be identical. While the statistical conditions for local graph stationarity in \cite{Girault2017,Girault2017_2} are defined in terms of the local graph power spectral density, we consider a milder definition of local stationarity in the vertex domain based on the work in \cite{Dahlhaus1996}, whereby the vertex signals in a neighborhood $\mathcal{V}_{m}$ exhibit the same first- and second-order moments, that is
\begin{align}
	\mu_{m} = \expect{x_{n}}, \quad \sigma_{m}^{2} = \var{x_{n}}, \quad \forall n \in \mathcal{V}_{m}.
\end{align}
In addition, we allow for a non-zero correlation between vertex signals in this neighbourhood, that is
\begin{align}
	\corr{x_{n},x_{k}} = \begin{cases}
		\rho_{m}, & n \neq k, \\
		1, & n=k,
		\end{cases} \quad \forall n,k \in \mathcal{V}_{m}.
\end{align}


\subsection{Doubly stochastic graph shift operators}

\label{section:DSGSO}

We next consider a class of doubly stochastic GSOs, denoted by $\S \in \domR^{N \times N}$, which exhibit the following properties
\begin{align} \label{eq:doubly_stochastic_properties}
	S_{mn} \geq 0, \quad \S\1 = \1, \quad \S^{\Trans}\1=\1
\end{align}
that is, $\S$ is a square, non-negative, matrix with columns and rows which sum up to unity. 


\begin{remark} \label{remark:Markov}
	The left-stochasticity property ($\S^{\Trans}\1=\1$) allows for a Markov (diffusion) matrix interpretation of the shift operator, whereby the $(m,n)$-th entry, $S_{mn}$, can be thought of as the transition probability of a random walker going from a vertex $n$ to a vertex $m$. Intuitively, the probability of going from a vertex $n$ to any vertex $m \in \mathcal{V}$ is equal to unity, i.e. $\sum_{m=1}^{N} S_{mn}=1$ (the $n$-th column sums up to unity).
\end{remark}



%

\begin{remark} \label{remark:norm_bound}
	The $L_{p}$-norm of a matrix, $\A$, is defined as
	\begin{align}
		\|\A\|_{p} = \begin{cases}
			\max_{j} \sum_{i}|A_{ij}|, & p=1,\\
			\sqrt{\lambda_{max}(\A^{\Trans}\A)}, & p=2,\\
			\max_{i} \sum_{j}|A_{ij}|, & p=\infty.
		\end{cases}
	\end{align}
	Since the largest eigenvalue of a doubly stochastic matrix is equal to unity \cite{Bapat1997}, and the rows and columns sum up to unity as well, then $\|\S\|_{p}=1$ for all $p=1,2,\infty$. Together with the Cauchy-Schwarz inequality, we obtain the following result
	\begin{align}
		\|\S\x\|_{p} \leq \|\S\|_{p}\|\x\|_{p} = \|\x\|_{p}, \quad \forall p=1,2,\infty. \label{eq:L2_norm}
	\end{align}
\end{remark}

\begin{remark} \label{remark:mean-preservation}
	The doubly stochastic shift preserves the mean of the graph signal values. To see this, consider a graph signal, $\x \in \domR^{N}$, with mean equal to $\mu$, that is, $\x = \mu \1 + \v$, where $\v \in \domR^{N}$ is a zero-mean signal, i.e. $\1^{\Trans}\v = 0$. Next, consider the doubly stochastic shift applied to the graph signal, to yield
	\begin{align}
		\y = \S\x = \S(\mu\1 + \v) = \mu\S\1 + \S\v = \mu\1 + \S\v
	\end{align}
	The mean of $\y$ is also $\mu$, since $\1^{\Trans}\S\v=\1^{\Trans}\v=0$.
\end{remark}

\begin{remark}
	From Remark \ref{remark:mean-preservation}, the doubly stochastic shift exhibits the $L_{1}$-isometry for non-negative graph signals, since $\|\x\|_{1}=N\mu$. From the Birkhoff–von Neumann theorem \cite{Birkhoff1946}, a doubly stochastic matrix decomposes into a convex combination of $k$ permutation matrices \cite{Arnold1987}, i.e. $\S = \sum_{i=1}^{k}a_{i}\P_{i}$ with coefficients $0 \leq a_{i} \leq 1$, and $k \leq (N-1)^{2}+1$. Since permutation matrices exhibit the $L_{1}$-isometry, so too does the doubly stochastic GSO, i.e. $\|\S\x\|_{1} = \|\x\|_{1}$.
	
\end{remark}

\begin{remark}
	Intuitively, Remarks \ref{remark:norm_bound}--\ref{remark:mean-preservation} describe the behaviour of a doubly stochastic graph shift as a diffusion toward a uniform graph signal, since the noise component of a signal $\x = \mu\1 + \v$ diffuses over a graph shift, i.e. $\|\S\v\|_{p} \leq \|\v\|_{p}$, while the mean of the graph signal, $\mu$, is preserved over shifts. To see this, recall that the largest eigenvalue of $\S$ is equal to unity, while the remaining eigenvalues lie on the closed unit disk. Furthermore, the eigenvector associated to the unit eigenvalue is $\frac{1}{\sqrt{N}}\1$ \cite{Bapat1997}. As a result, we obtain the following convergence
	\begin{align}
		\lim_{k \to \infty} \S^{k}\x = \tfrac{1}{N}\1\1^{\Trans}\x = \mu\1
	\end{align}
	This describes the characteristic behaviour of a diffusion process asymptotically approaching a uniform signal.

\end{remark}

In practice, the actual probabilities of vertex transition are often unknown, however, these can be inferred using the available information of the graph domain topology, implied by the weight matrix, $\W$. The graph edges weights, $W_{mn}$, can be defined through domain knowledge, based on the geometry of vertex positions, or based on data similarity methods for learning the underlying graph topology \cite{Coifman2006,Eldar2017,Stankovic2019_2}. Once a weight matrix, $\W$, is defined, there exist several techniques for obtaining a doubly stochastic GSO, $\S$, with the properties in (\ref{eq:doubly_stochastic_properties}). The most well-known procedure is the \textit{Sinkhorn-Knopp} algorithm which can retrieve $\S$ through an alternating normalization of $\W$ by its row and column sums \cite{Sinkhorn1964,Sinkhorn1967,Knight2008}. The iterative procedure is summarised in Alg. \ref{alg:Sinkhorn-Knopp}, with the operator $\mathcal{D}: \domR^{N} \mapsto \domR^{N\times N}$ defined as $\mathcal{D}(\x)=\diag{\x}$.

\begin{algorithm}
	\caption{Sinkhorn-Knopp algorithm}\label{alg:Sinkhorn-Knopp}
	\begin{algorithmic}[1]
		\Procedure{Sinkhorn-Knopp}{$\W$}
		\State $\r \leftarrow \1$
		\While{not converged}
		\State $\c \leftarrow \mathcal{D}(\W^{\Trans}\r)^{-1}\1$
		\State $\r \leftarrow \mathcal{D}(\W\c)^{-1}\1$
		\EndWhile
		\State $\S \leftarrow \mathcal{D}(\r)\W\mathcal{D}(\c)$
		\State \textbf{return} $\S$
		\EndProcedure
	\end{algorithmic}
\end{algorithm}


Notice that in general $\W$ can be directed. This procedure is also known as the \textit{iterative proportional scaling algorithm} \cite{Stephan1940,Stephan1942}. Other techniques have also been proposed, based on the minimisation of $\|\W-\S\|$ \cite{Zass2005,Zass2006,Wang2016}.

Doubly stochastic normalizations of the weight matrix, $\W$, have already been employed to enhance graph-based clustering models \cite{Zass2006,Chang2009,Wang2010,Ding2011,Ding2012,Wang2016}, to improve the convergence rate of graph filter designs for large scale random networks \cite{Moura2013,Moura2017}, and for satellite communication systems \cite{Brualdi1988}. Instead, we next investigate an alternative application of the doubly stochastic matrix as a shift operator for random graph signals, and demonstrate its boundedness properties -- a prerequisite to its use in real world applications.


\section{Boundedness of doubly stochastic GSOs}

We begin by considering a doubly stochastic shift applied to the $m$-th vertex signal, $x_{m}$, to obtain the expression
\begin{align}
	\mathcal{S}(x_{m}) = \sum_{n \in \mathcal{V}_{m}} S_{mn}x_{n} \label{eq:shift}
\end{align} 
The variance of the shifted signal is then given by
\begin{align}
	\var{\mathcal{S}(x_{m})} = \expect{\mathcal{S}(x_{m})^{2}} - \expect{\mathcal{S}(x_{m})}^{2}
\end{align}
Upon rearranging the above equation, we obtain the expression for the \textit{expected} power of the shifted random graph signal
\begin{align}
	\expect{\mathcal{S}(x_{m})^{2}} = \expect{\mathcal{S}(x_{m})}^{2} + \var{\mathcal{S}(x_{m})} \label{eq:boundedness_master}
\end{align}
In this way, the examination of the boundedness properties over graph shifts reduces to performing a \textit{statistical consistency} analysis of the graph shift operator.

Given the difficulty of evaluation of the statistical consistency of the graph shift for an arbitrary random graph signal, we consider a locally stationary graph signal as described in Section \ref{sec:random_graphs}, whereby the signal at a vertex $n \in \mathcal{V}_{m}$ is assumed to be distributed according to $x_{n} \sim \mathcal{N}(\mu,\sigma^{2})$, and the correlation between vertex signals in this neighbourhood is assumed to be $\rho = \corr{x_{n},x_{k}}$ for all $n,k \in \mathcal{V}_{m}$, $n \neq k$.

\subsection{Bias}

The doubly stochastic shift in (\ref{eq:shift}) is an \textit{unbiased} estimator of the mean, since each of its rows sums up to unity, that is
\begin{align}
	\expect{\mathcal{S}(x_{m})} & = \sum_{n \in \mathcal{V}_{m}} \! S_{mn} \expect{x_{n}} = \mu \! \sum_{n \in \mathcal{V}_{m}} \! S_{mn} =  \mu \label{eq:bias}
\end{align}

\subsection{Asymptotic consistency}

\label{sec:consistency}

Consider the variance of the graph shift in (\ref{eq:shift}), given by
\begin{align}
	\var{\mathcal{S}(x_{m})} & \! = \! \sum_{n \in \mathcal{V}_{m}} \! \sum_{k \in \mathcal{V}_{m}} \!\! S_{mn}S_{mk} \, \cov{x_{n},x_{k}} \notag\\
	&  =  \sigma^{2} \biggl( \sum_{n \in \mathcal{V}_{m}} \!\! S_{mn}^{2} + \rho \! \sum_{\subalign{n & \in \mathcal{V}_{m}\\ n & \neq k}} \! \sum_{\subalign{k & \in \mathcal{V}_{m}\\ k & \neq n}} \!\! S_{mn}S_{mk} \biggr) \label{eq:GRW_variance}
\end{align} 
The Cauchy-Schwarz inequality can be employed to assert the following bounds on the last term in the above expression
\begin{align}
	 \sum_{\subalign{n & \in \mathcal{V}_{m}\\ n & \neq k}} \! \sum_{\subalign{k & \in \mathcal{V}_{m}\\ k & \neq n}} \!\! S_{mn}S_{mk} & \leq \biggl( \sum_{n \in \mathcal{V}_{m}} S_{mn} \biggr)^{2} \!\! \leq  N_{m} \! \sum_{n \in \mathcal{V}_{m}} \!\! S_{mn}^{2}
\end{align}
Therefore, the variance is bounded from above according to
\begin{align} \label{eq:GRW_variance2}
	\var{\mathcal{S}(x_{m})} &  \leq  \sigma^{2}\left(1 + N_{m} \rho\right) \! \sum_{n \in \mathcal{V}_{m}} \!\! S_{mn}^{2}
\end{align}
Furthermore, an upper bound to the term $\sum_{n \in \mathcal{V}_{m}} S_{mn}^{2}$ can be obtained from the Kantorovich inequality \cite{Kantorovich1948}, which yields
\begin{align}
 \biggl(  \sum_{n \in \mathcal{V}_{m}} \!\! S_{mn}^{2}  \biggr)  \biggl(  \sum_{n \in \mathcal{V}_{m}} \!\! 1^{2} \biggr)  \leq  \frac{(L +  U)^{2}}{4LU} \biggl(  \sum_{n \in \mathcal{V}_{m}} \!\! S_{mn}  \biggr)^{\!\!2}
\end{align}
where $L$ and $U$ denote respectively the lower and upper bounds on the possible values of $S_{mn}$, which in our case are
\begin{align}
	0 < L \leq S_{mn} \leq U < 1, \quad \forall m,n
\end{align}
We finally obtain the following result
\begin{align} \label{eq:Kantorovich}
\sum_{n \in \mathcal{V}_{m}} S_{mn}^{2} \leq \frac{1}{N_{m}} \frac{(L+U)^{2}}{4LU} < 1
\end{align}

\begin{remark}\label{remark:lower_upper_bounds}
	The lower and upper bounds of $S_{mn}$, $L$ and $U$, are invariant to the neighbourhood size, $N_{m}$, as they are determined solely from the underlying physics of the problem.
\end{remark}

From Remark \ref{remark:lower_upper_bounds}, with an increase in the incoming neighbour size, $N_{m}$, in the limit we obtain the following upper bound on the graph shift variance, since from (\ref{eq:GRW_variance2}) and (\ref{eq:Kantorovich})
\begin{align}
\lim_{N_{m} \to \infty}  \var{\mathcal{S}(x_{m})} & \leq  \lim_{N_{m} \to \infty}  \sigma^{2} \left(1 + N_{m} \rho\right) \! \sum_{n \in \mathcal{V}_{m}}  S_{mn}^{2} \notag\\
& \leq  \lim_{N_{m} \to \infty}  \frac{\sigma^{2} \left(1 + N_{m} \rho\right)}{N_{m}} \frac{(L+U)^{2}}{4LU} \notag\\
& = \rho \sigma^{2} \frac{(L+U)^{2}}{4LU}\label{eq:asymptotic_consistency}
\end{align} 
This proves that, as desired, the degree of statistical inconsistency of the doubly stochastic graph shift is upper bounded.

\begin{remark}
	For an \textit{i.i.d.} random graph signal ($\rho = 0$), the upper bound on the shift variance vanishes, since from (\ref{eq:asymptotic_consistency})
	\begin{align}
		\lim_{N_{n} \to \infty}  \var{\mathcal{S}(x_{m})} = 0
	\end{align} 
	Therefore, the doubly stochastic graph shift is \textit{statistically consistent} for \textit{i.i.d.} graph signals.
\end{remark}

\subsection{$L_{2}$-norm upper boundedness}

An upper $L_{2}$-boundedness of the shift operator can also be proven for an asymptotic increase in the neighbourhood size, $N_{m}$. Starting from (\ref{eq:boundedness_master})-(\ref{eq:bias}) and by employing the inequality in (\ref{eq:asymptotic_consistency}), we obtain the following asymptotic behaviour
\begin{align} \label{eq:shift_norm_lower_bound}
	\lim_{N_{m} \to \infty} \expect{\mathcal{S}(x_{m})^{2}} & \leq \mu^{2} + \rho \sigma^{2} \frac{(L+U)^{2}}{4LU} 
\end{align}
which proves the asymptotic $L_{2}$-norm upper boundedness of the doubly stochastic graph shifted signal.

\begin{remark}
	Notice that the bias term in (\ref{eq:shift_norm_lower_bound}), given by
	\begin{align}
	\frac{(L + U)^{2}}{4LU} = \left(\frac{\frac{1}{2}(L+U)}{\sqrt{LU}}\right)^{2} \geq 1
	\end{align}
	is simply the square of the ratio of the arithmetic mean to the geometric mean of the bounds $L$ and $U$. The AM-GM inequality therefore asserts that this bias term is bounded by unity from below, with the equality attained for $L=U$. In other words, the magnitude of this bias term is minimised by maximising the ratio of the bounds, $\frac{L}{U}$, or equivalently, by promoting the homogeneity of the graph edge weights, $W_{mn}$, within each neighbourhood, $\mathcal{V}_{m}$. This suggests that the design of neighbourhoods, or the location of additional vertices, may be chosen so as to maximise the ratio, $\frac{L}{U}$, thus in turn tightening the upper boundedness of the GSO.
\end{remark}

\subsection{$L_{2}$-norm lower boundedness}

\label{sec:L2}

Since the variance of the graph shifted signal is strictly non-negative, $\var{\mathcal{S}(x_{m})} \geq 0$, from (\ref{eq:boundedness_master})-(\ref{eq:bias}) we obtain the lower bound of the $L_{2}$-norm of the shifted signal in the form
\begin{align}
	\expect{\mathcal{S}(x_{m})^{2}} \geq  \mu^{2}
\end{align} 
which is a direct consequence of Jensen's inequality.

\begin{remark}
	The shift operator is asymptotically $L_{2}$-norm isometric for \textit{i.i.d.} graph signals, since for $\rho = 0$ the lower and upper bounds of $\lim_{N_{m} \to \infty} \expect{\mathcal{S}(x_{m})^{2}}$ coincide, leading to the following desired result
	\begin{align}
		\lim_{N_{m} \to \infty} \expect{\mathcal{S}(x_{m})^{2}} = \mu^{2}
	\end{align}
	Therefore, the doubly stochastic GSO preserves the power of the mean for \textit{i.i.d.} graph signals.
\end{remark}

\subsection{Boundedness of systems on a graph}

For a random graph input signal, a linear system of order $K$ is defined as \cite{Sandryhaila2013}  \vspace{-0.2cm}
\begin{align}
	\y = \sum_{k=0}^{K} h_{k} \S^{k}\x \label{eq:system}
\end{align}
where $h_{k}$ are the system coefficients. Based on the above derived boundedness properties, the class of systems based on the doubly stochastic shift also exhibits desirable boundedness properties. Starting from (\ref{eq:L2_norm}), we obtain the inequality $\|\S^{k}\x\|_{p} \leq \|\x\|_{p}$ for $p=1,2,\infty$ and all $k\geq0$. Together with Minkowski's inequality, we can show that the graph system output, $\y$, is $L_{2}$-norm upper bounded as follows
\begin{align}
\|\y\|_{p} & \leq \sum_{k=0}^{K} |h_{k}| \| \S^{k}\x\|_{p} \leq  \sum_{k=0}^{K} |h_{k}| \|\x\|_{p}, \quad \forall p=1,2,\infty.
\end{align}


\section{Numerical Example}



\begin{figure}[ht]
	\vspace{-0.2cm}
	\centering
	\begin{subfigure}[t]{0.24\textwidth}
		\centering
		\includegraphics[width=0.9\textwidth, trim={6cm 13.7cm 0.5cm 3cm}, clip]{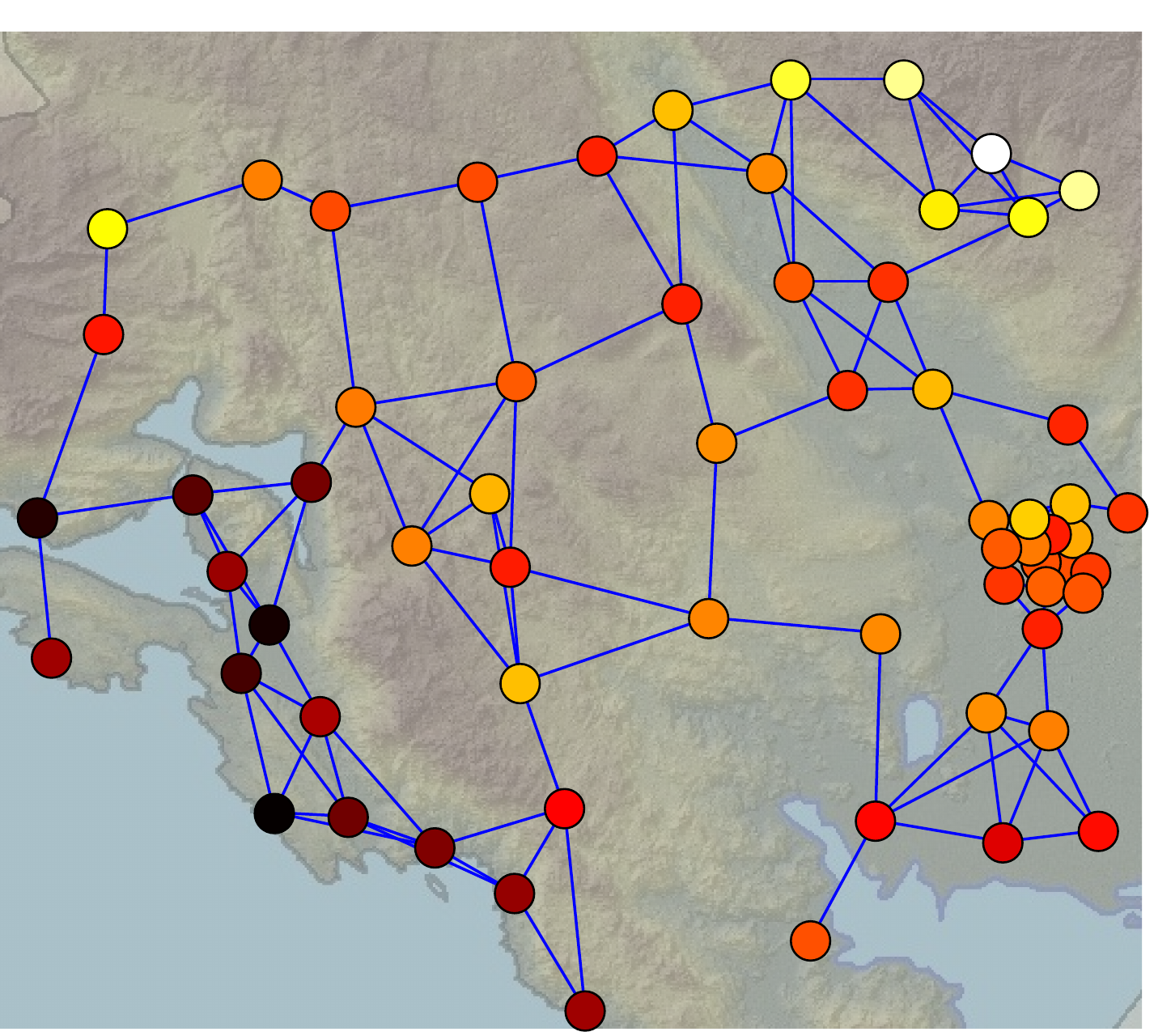} 
		\caption{{\small  Observed temperature field}}  
		{\small ($\text{SNR} = 14.0 \, \text{dB}$)}  
		\label{fig:orig}
	\end{subfigure}
	\hfill
	\begin{subfigure}[t]{0.24\textwidth}   
		\centering 
		\includegraphics[width=0.9\textwidth, trim={6cm 13.7cm 0.5cm 3cm}, clip]{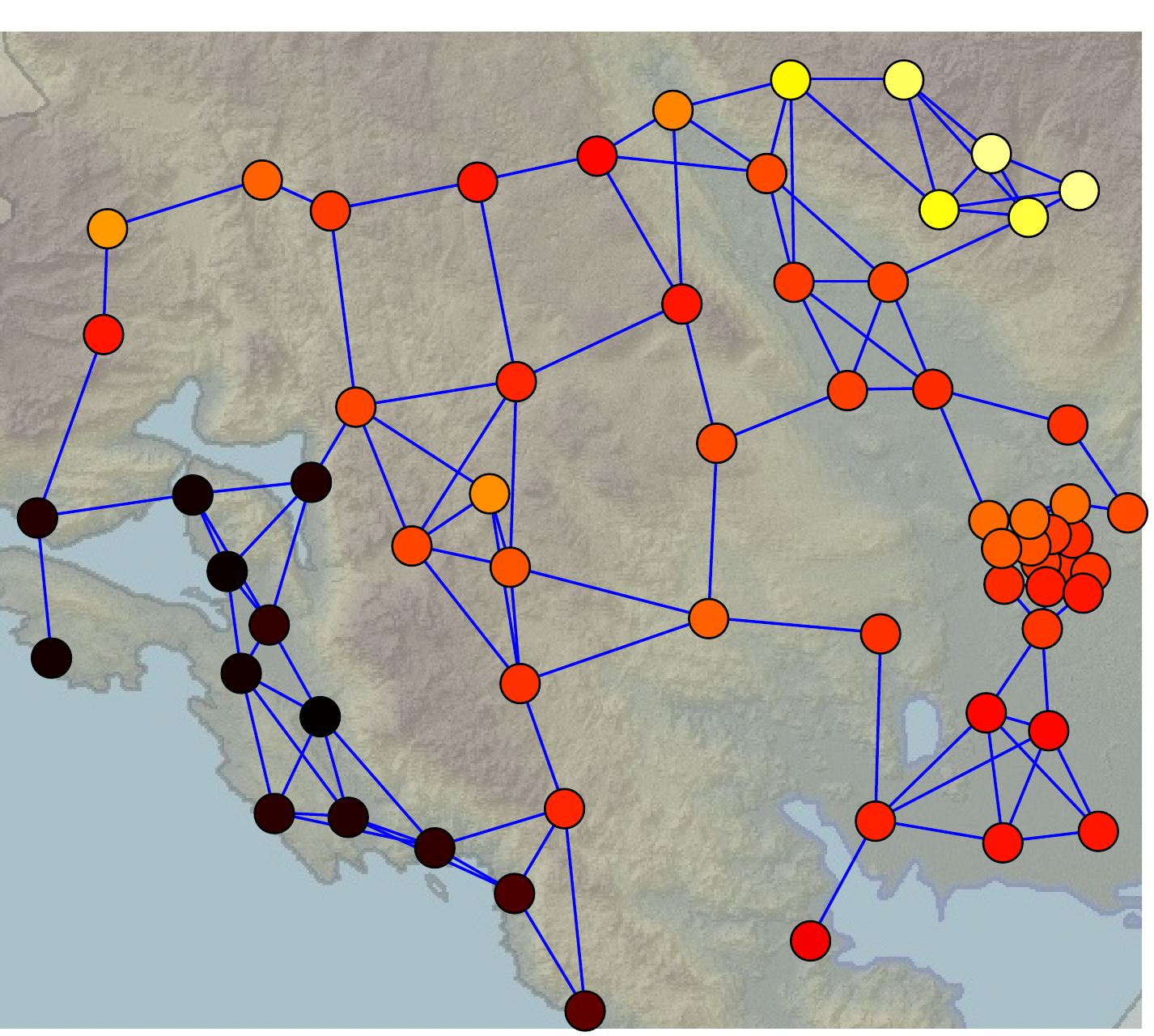} 
		\caption[]%
		{{\small GSO expectation.}} 
		{\small ($\text{SNR} = 19.8 \, \text{dB}$)}  
		\label{fig:GRW}
	\end{subfigure}
\end{figure}

Consider a multi-sensor setup, described in \cite{Stankovic2019_2}, for measuring a temperature field in a geographic region. This temperature field consists of $N=64$ sensor measurements in total, as shown in Figure \ref{fig:orig}). Each measured sensor signal was corrupted with synthetic Gaussian noise to emulate the possible adverse effects of the local environment on sensor readings or faulty sensor activity. The $n$-th vertex can therefore be mathematically expressed as $x_{n} \sim \mathcal{N}(\mu_{n},\sigma^{2})$, where $\mu_{n}$ is the true temperature at the $n$-th sensor (vertex). In our study, the standard deviation of the noise was set to $\sigma=2$, to yield the signal-to-noise ratio in $x_{n}$ of $\text{SNR} = 14.0 \, \text{dB}$.

The doubly stochastic GSO was employed as an expectation operator to estimate the true temperature from the observed temperature field. The weight matrix entries, $W_{mn}=e^{-r_{mn}^{2}}$, were specified based on the geographical distance between vertices, $r_{mn}$, thereby accounting for the difference in latitude, longitude and altitude. The matrix $\S$ was obtained from $\W$ using the Sinkhorn-Knopp algorithm described in Alg. \ref{alg:Sinkhorn-Knopp}. The denoised temperature field is illustrated in Figure \ref{fig:GRW}), whereby the shifted graph signal attained a $5.8 \, \text{dB}$ SNR gain.

\section{Conclusions}

The boundedness of a class of doubly stochastic graph shift operators (GSOs) has been established by performing a statistical consistency analysis of the graph shift. This has been achieved based on the dual role of the doubly stochastic GSO as a Markov (diffusion) matrix and as an unbiased graph expectation operator. The usefulness of doubly stochastic GSOs for operating on random graph signals has been demonstrated through analysis, and a practical real-world multi-sensor temperature estimation example.

\pagebreak

\footnotesize

\bibliographystyle{IEEEtran}
\bibliography{./Bibliography} 

\end{document}